# ON THE PROCEDURAL STRUCTURE
# OF LEARNING ECOSYSTEM
# TOWARD COMPETENCY LEARNING MODEL

NGUYEN MANH HUNG[*], NGUYEN HOAI NAM[*]

**ABSTRACT**

*Learning Ecosystem is new model for learning, that addresses to holistic learning model with attention to practical implementation. This paper is conducting the further study on detailed structure of learning ecosystem in component and procedural view. As case study, it connects Learning Ecosystem to Competency Education as Competency Learning Ecosystem Model for reference for practical use.*

**TÓM TẮT**

***Cấu trúc thủ tục của hệ sinh thái học tập theo mô hình học tập phát triển năng lực***

*Mô hình hệ sinh thái học tập là một hình học tập chính thức, mới và đang được quan tâm nghiên cứu triển khai. Bài báo tiến hành nghiên cứu chi tiết hệ sinh thái học tập theo các thành phần cấu trúc và quy trình. Trong phần vận dụng, chúng tôi đề xuất mô hình hệ sinh thái học tập theo hướng phát huy năng lực người học.*

***Từ khóa*:** hệ sinh thái học tập, thiết kế học tập, ngôn ngữ mẫu, lí thuyết kết nối, giáo dục năng lực, thành tố học tập.

## 1. Introduction

Learning Ecosystem concept and model have been introduced and described by Nguyen Manh Hung [9], [10] as new model for learning system with an attention to holistic view and ability to practical implementation. This is extension of traditional models of learning, which have been relied on closed structure and space as classroom learning space and teacher-student learning hierarchy so far. The examples of these traditional learning models are Dunn & Dunn Learning Styles Model [2], [3], and Kolb's Experiential Learning Model [6]. These learning models facilitate the process of linking instructional activities to individual learning styles, thereby increasing the learner's ability to acquire and retain knowledge. This is typical for traditional learning models that focus on student as his/her personality without the connection to learning

_____________________________________________________________________________
[*] Ph.D., Hanoi National University of Education, Hanoi, Vietnam

ecology, or focus on instruction perspective of learning as content-based learning models.

Based on connectivism [13], Siemens has been proposed learning model with more broad space and structure of learning and characterized by such principles as openness, diversity, autonomy and interactivity/connectivity and shown in Figure 1 [14].

In order to make this Learning Ecology having capability of practical application, the Learning Ecosystem concept is defined as connective together systems of learning subjects, learning contents, learning contexts and learning technologies as Figure 2 [9].

Next part of this paper is reserved for description of coherent structure of Learning Ecosystem with defining components, entities of each system and their close relative elements. Third part presents one case study for connecting Competency Education to Learning Ecosystem as practical application of this model.

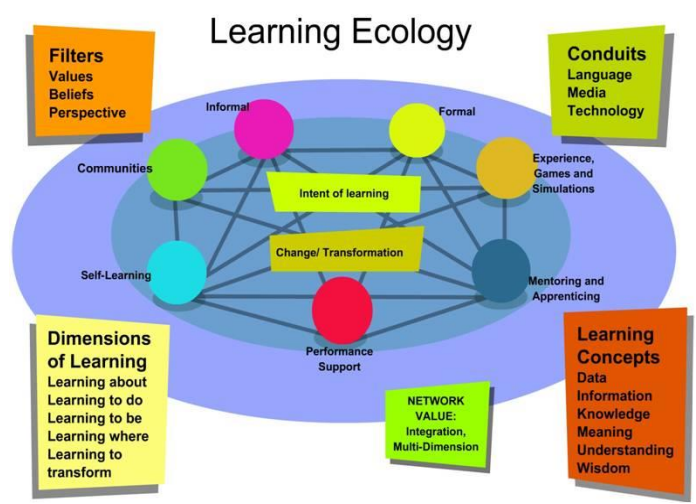

***Figure 1.*** *Learning Ecology Model*

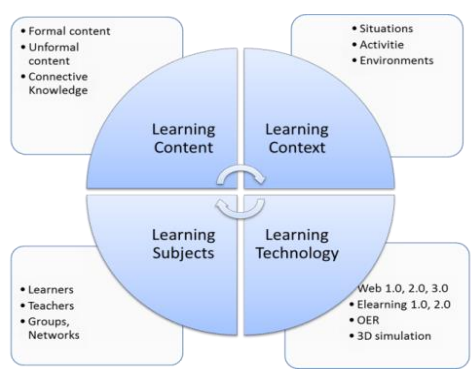

***Figure 2.*** *Learning Ecosystem Elements*

**2.     Structure of Learning Ecosystem**

In this part we list out main components of Learning Ecosystem and theoretical base elements for designing/defining/guiding these components, for instance, learning content is designed by learning design methods like IMS LD (IMS Global Learning Consortium Learning Design) or learning context is designed by context methods like pattern language.

Learning Ecosystem's and underground components are shown in Figure 3.

*2.1.  Learning Content System*

Learning Content has been playing key role in any education system. For creating learning content there are huge interests and forces from educators in various research fields such as instructional design, learning design (LD), learning object (LO). Building of learning curricula, textbook system is a heart of any education reform. Thus, for example, Vietnam have renewed textbook system in 2000-2005 for education reform of a period of 2000-2010, and now this is new textbook system will be designed for reform of a period from 2015.

Beside the formal contents such as textbook system (textbooks, exercises books, complements) or lesson's instruction and curricula, there are other informal contents for learning such as wikis, OER (Open Education Resource) or WebPages, which are becoming more and more important for learning nowadays. In terms of online learning resources, there are LOs sharable and open in the internet and being packed by open standard and metadata (e.g. ADL SCORM (Advanced Distributed Learning SCORM=Sharable Content Object Reference Model)) and managed by various Learning Management Systems over the internet.

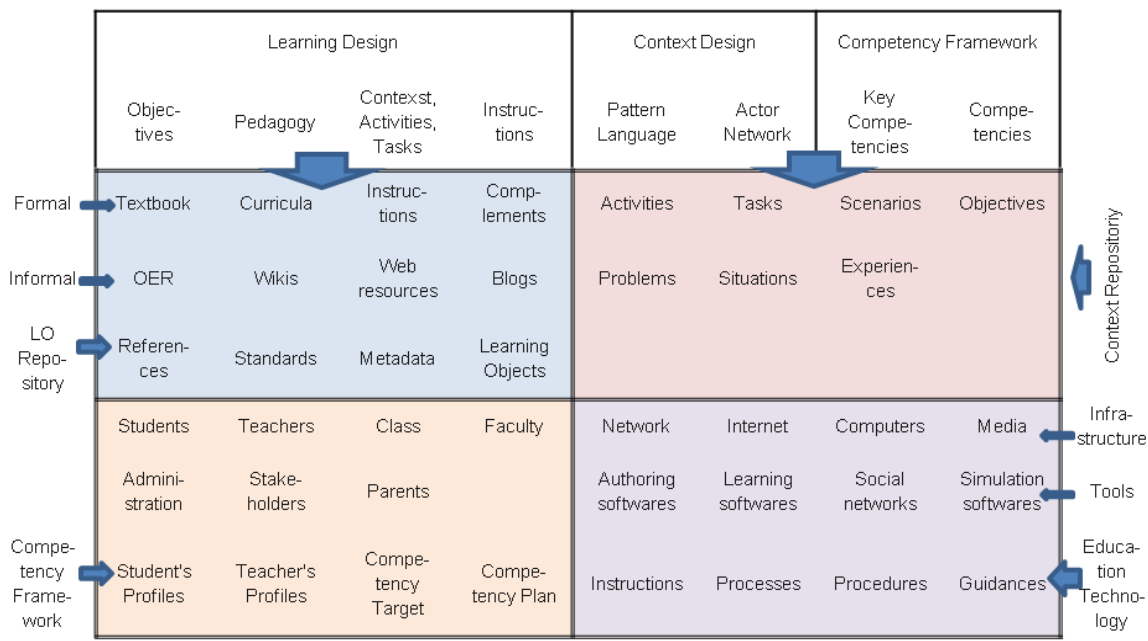

*Figure 3. Extented Structure of Learning Ecosystem*

There are various instructional design methods and models for creation and design learning content:

- Merrill's First Principles of Instruction (http://id2.usu.edu/Papers/5FirstPrinciples.PDF).
- ADDIE Model (http://en.wikipedia.org/wiki/ADDIE_Model).
- Dick and Carey Model (http://en.wikipedia.org/wiki/Instructional_design#Dick_and_Carey).
- Kemp's Instructional Design Model (http://edutechwiki.unige.ch/en/Kemp_design_model%20).
- Gagné's Nine Events of Instruction (http://edutechwiki.unige.ch/en/Nine_events_of_instruction).
- Bloom's Learning Taxonomy (http://ww2.odu.edu/educ/roverbau/Bloom/blooms_taxonomy.htm)
- Kirkpatrick's 4 Levels of Training Evaluation (http://www.businessballs.com/kirkpatricklearningevaluationmodel.htm).

Among various LD methods and approaches, IMS LD [5] becomes most popular standard for LD and application in practice, based on Rob Koper Educational Modeling Language (EML). [7]

*Table 1. IMS LD elements*

| **Management** | Interoperability parameters with LMS |
|---|---|
| **Pedagogical/Instructional** | Pedagogical Information |

| **Activity/Task** | Educative processes and activities. Collaborative tasks and activities |
|---|---|
| **Sequencing** | Sequencing, prerequisites, deadlines, dependencies |
| **Structure** | Navigational model |
| **Content** | Small LO's, assets and formatted content |

As is shown in Table 1, IMS LD has some elements similar to components of context system such as activities, tasks, objectives. Hence, while LD using pedagogy, activities, tasks or contexts for designing learning contents, the context system of Learning Ecosystem is using these elements connectively for choosing or defining appropriate content from these listed components and defining learning subjects and learning technology too.

There is the difference between learning designing of content as LD approaches and methods for content's authoring in one side and choosing or defining appropriate learning content component (e.g. textbook or wikis) based on learning context in another side. In this Learning Ecosystem model this is uplift from authoring or creating content. This work is reserved for educators and teachers.

Learning Process from the view to Learning Content System:

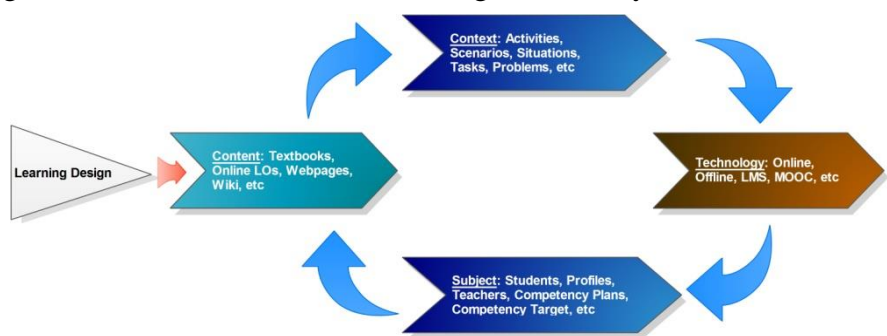

*Figure 4. Learning Content Procedure*

## 2.2. *Learning Context System*

In Learning Ecosystem model, this Learning Context System plays a key role, and it consists of such components as activities, problems, tasks, situations, experiences, objectives, etc and can be designed by various context designing methods or approaches, like Pattern Language and Actor Network Theory. Learning context is defined and designed based on many factors such as learning filters, learning dimensions, intent of learning, etc [14], but here we have limited to competency framework only (key competencies, competency level, competency target).

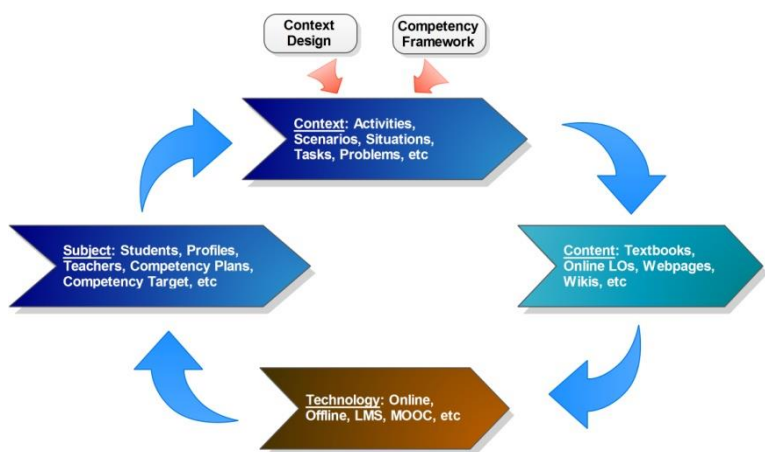

*Figure 5. Learning Context Procedure*

## 2.3. Learning Subjects System

This system can be limited to learners and teachers or expanded to relevant people such as administration, parents or stakeholders. There are included student profiles, teacher profiles, and student learning plan or competency target for matching with current student profiles in order to define the gap to appropriate competency requirement as per targeted to student. At this era of internet and social networks, emerging learning communities, networks or groups play more greater role for personal learning today. It also emphasizes the importance of peer learning and participation of a learner as the provider of learning content beside a teacher in learning process.

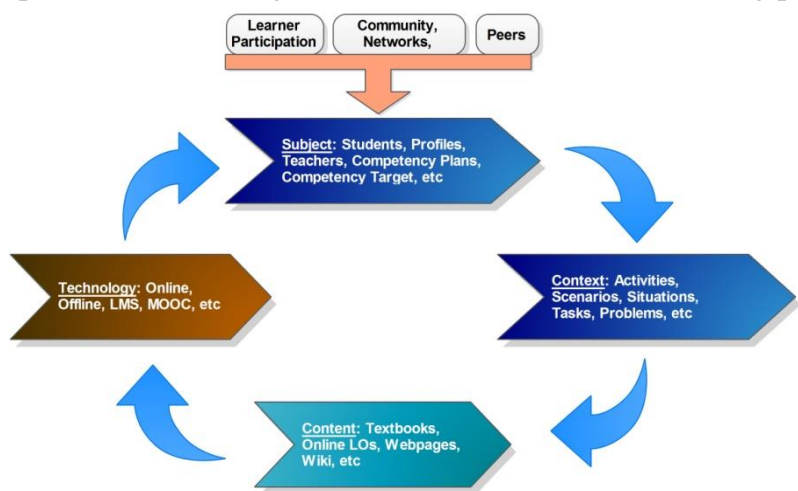

*Figure 6. Learning Subjects Procedure*

## 2.4. Learning Technology System

We are exposed that ICT, particularly internet and WWW has been impacting powerfully to education. Hence there are a lot of informal learning materials and methods based on internet through huge web learning resources such as wikis, LMSs (Learning Management System), OERs, MOOCs (Massive Open Online Course), social networks. Supporting to formal learning, there are various software applications such as simulations/games, mind map softwares, mathematical softwares. Supporting to teaching, there are various authoring tools, education softwares and media for delivering teaching pedagogy to students. In term of learning management, there are various tools and softwares, standards, frameworks for quality assessment and process management.

The supportive Education Technology is specialized field of education research that regards creating, designing or using these above mentioned tools, softwares, processes and procedures. Inside the bound of this learning technology system, we can combine appropriate technology components in connection with context, content and subjects to form a technology base for learning process.

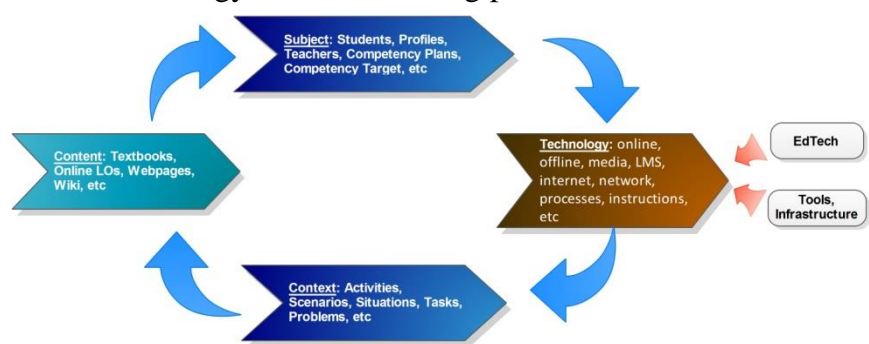

*Figure 7. Learning Technology Procedure*

### 3. Case Study: Competency Learning Ecosystem

In natural ecology, each living entity has to survive his/her life cycle in surrounding environment. For self-development, living entity like animal is required to being capable of acquiring living skills and experiences or living competencies. These competencies can be owned genetically or learned through active connectivity with environment. Similarly, in human society, every person has to have appropriate competency set to meet living and working demands and requirements. The competency education approach has been adopted largely over the world and now is starting formation in our country. This competency education (or competence education) is built based on competency framework/standard that normally consists of key competencies (e.g. USA Common Core State Standards or European Reference Framework for Lifelong Education [1], [11], and being implemented through various systems, methods or approaches. For avoiding misunderstandings between terms "competence" and "competency" [8], we are referring to competency terminology for whole this paper.

Competency Education is rooted in the notion that education is about mastering a set of skills and knowledge, not just moving through a curriculum. In competency education, students keep working on specific skills or knowledge until they can demonstrate their understanding and ability to apply them; they then move to the next material while continuing to use what they have already learned. Students cannot advance simply by showing up to class on a sufficient number of days and earning a grade just above failing. Instead they must meet standards (also known as competencies, performance objectives, or learning targets) at a predetermined level of proficiency [12].

Using Learning Ecosystem model, we would like to present one case study for competency learning ecosystem, that addresses to competency learning model.

The process of establishing and developing competencies for students is illustrated in Figure 8.

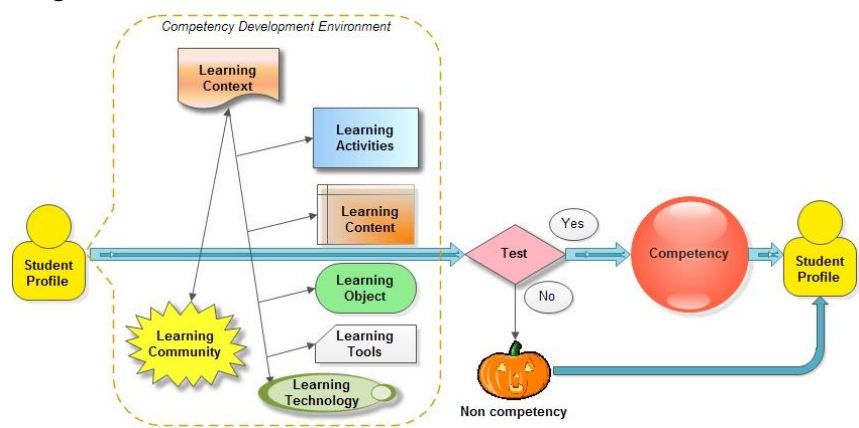

*Figure 8. Learning Competency Pathway through Learning Ecosystem*

The competency education is learner-centric, personalizing learner progress. Hence, the learning process takes a place over the path to meet competency target. In order to master one given competency, the student takes part in Learning Context as environment, circumstance or situation of learning. This learning context from its own will define Learning Activities, in which the student will be involved for acquiring appropriate knowledge and skill set as parts of Learning Content. The student is attaining given competencies in connection with a set of Learning Subjects such as community, teachers, peer-learners, stakeholders and through the Learning Technology that supports to the student.

In the competency-based education model, the assessment of competency for students is crucial key element for competency progression and attainment of students. A feedback from these assessments is resulted in a student profile, which is mapped with competency target for defining personalized activities within appropriate learning contexts. This learning cycle can be executed through Competency Development Environment that extracted from Learning Ecosystem as shown in Figure 8.

From the view of competency education system, this competency assessment and attainment cycle is shown in Figure 9 [4],

In connection with Learning Ecosystem, this process is shown at more detailed view in Figure 10.

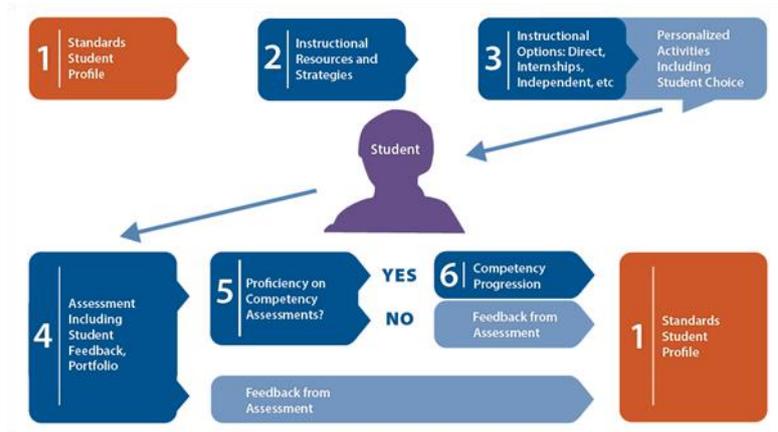

*Figure 9. Competency-Based Learning Cycle*

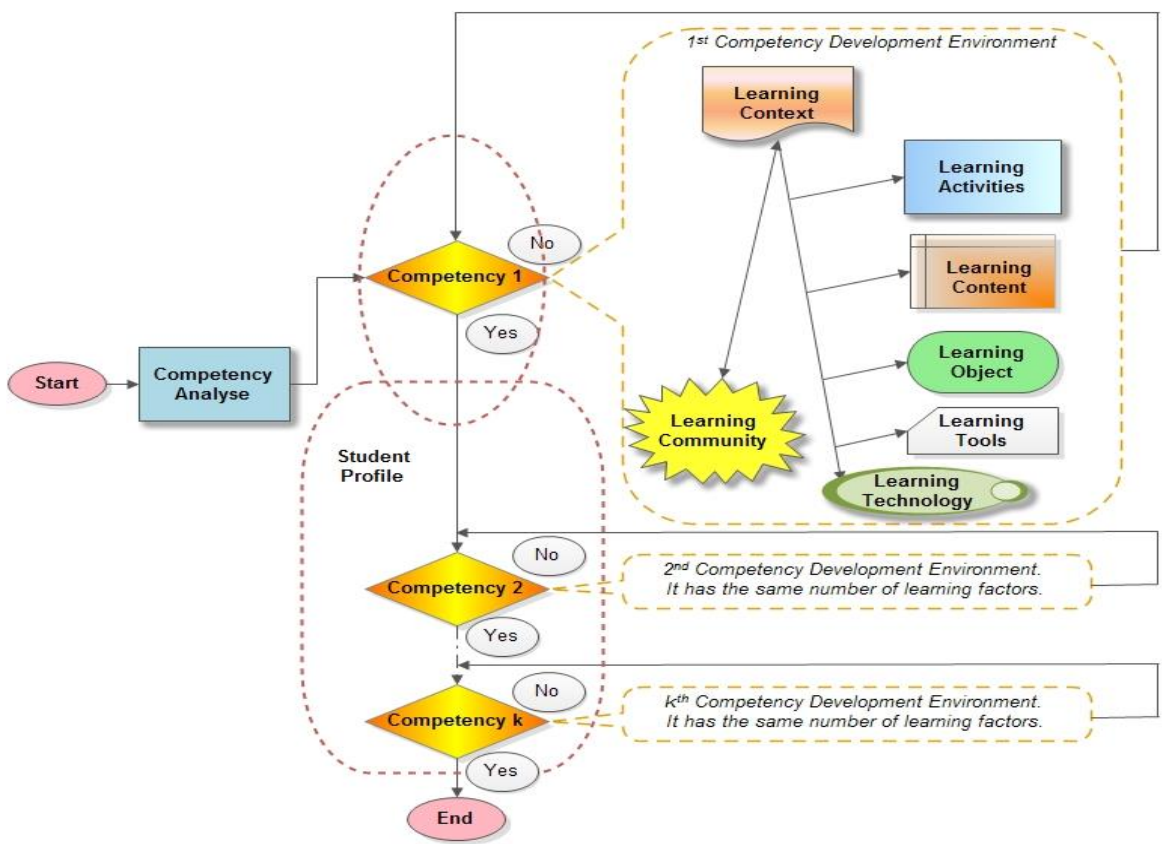

*Figure 10. Competency Attainment Process in Competency Learning Ecosystem*

**4.**   **Conclusion**

In this paper, we have been described details of structure of Learning Ecosystem Model for showing clear picture of this model that can be used for an implementation of learning environment updated to technology and social transformation nowadays. This Learning Ecosystem is providing holistic view to learning and teaching practice and showing in one case study for Competency Education. We hope that proposed Learning Ecosystem model will be usefull to Education Reform Program for our country as new view and approach for designing key elements of this Program.